\documentclass[twocolumn,showpacs,amssymb,prl,aps]{revtex4}
\usepackage{graphicx}
\usepackage{bm}

\begin{document}

\title{Nanoscale magnetic structure of ferromagnet/antiferromagnet manganite multilayers}

\author{D. Niebieskikwiat}
\affiliation{Department of Physics, University of Illinois at
Urbana-Champaign, Urbana, Illinois 61801, USA}

\author{L.E. Hueso}
\affiliation{Department of Materials Science, University of
Cambridge, Cambridge CB2 3QZ, UK}

\author{J.A. Borchers}
\affiliation{NIST Center for Neutron Research, National Institute
of Standards and Technology, Gaithersburg, Maryland 20899, USA}

\author{N.D. Mathur}
\affiliation{Department of Materials Science, University of
Cambridge, Cambridge CB2 3QZ, UK}

\author{M.B. Salamon}
\affiliation{Department of Physics, University of Illinois at
Urbana-Champaign, Urbana, Illinois 61801, USA}

\begin{abstract}

Polarized Neutron Reflectometry and magnetometry measurements
have been used to obtain a comprehensive picture of the magnetic
structure of a series of
La$_{2/3}$Sr$_{1/3}$MnO$_{3}$/Pr$_{2/3}$Ca$_{1/3}$MnO$_{3}$
(LSMO/PCMO) superlattices, with varying thickness of the
antiferromagnetic (AFM) PCMO layers ($0\leq t_A \leq 7.6$ nm).
While LSMO presents a few magnetically frustrated monolayers at
the interfaces with PCMO, in the latter a magnetic contribution
due to FM inclusions within the AFM matrix was found to be
maximized at $t_A\sim 3$ nm. This enhancement of the FM moment
occurs at the matching between layer thickness and cluster size,
where the FM clusters would find the optimal strain conditions to
be accommodated within the ``non-FM" material. These results have
important implications for tuning phase separation via the
explicit control of strain.

\end{abstract}


\maketitle

Nanostructured magnetic materials are already finding
applications in magnetic recording and magnetic memory devices.
One important focus has been on tunneling magnetoresistance (TMR),
in which ferromagnetic (FM) layers having spin-polarized
conduction electrons are separated by a non-magnetic insulating
barrier \cite{ziese02,TMR}. Double-exchange magnets such as
La$_{2/3}$Sr$_{1/3}$MnO$_{3}$ and La$_{2/3}$Ca$_{1/3}$MnO$_{3}$,
because of their high degree of spin polarization, have been
considered as the metallic layers along with various insulating
barriers such as SrTiO$_{3}$ (STO), LaAlO$_{3}$, and NdGaO$_{3}$.
Results have not been encouraging. One possible problem with
these systems lies in the interfacial magnetization, which may
fall off more rapidly with increasing temperature than does the
bulk magnetization \cite{TMR}.

As a different approach to TMR devices, we have fabricated
all-manganite multilayers utilizing antiferromagnetic (AFM)
insulating manganites as the barrier
\cite{cheng98,li02,krivorotov01,jo03}. Our goal is to study the
magnetic properties of such multilayers with special attention to
two different aspects. First, we expect that the similarities in
lattice structure and stoichiometry at the manganite-manganite
interfaces will minimize the suppression of interfacial
magnetization. Second, we explore the possibility of inducing
magnetization in the AFM spacer and thereby having a
magnetic-field-sensitive tunneling barrier, which would contribute
to the TMR effect. AFM materials with FM instabilities, i.e.
phase-separated manganites, are thus ideal candidates for the
insulating layers \cite{dagotto01}.

We have chosen as the barrier layer Pr$_{2/3}$Ca$_{1/3}$MnO$_{3}$
(PCMO), which is AFM and insulating, but supports nanoscale FM
droplets within the AFM matrix
\cite{smolyaninova02,radaelli01,saurel06}. Because of its high
Curie temperature, we use La$_{2/3}$Sr$_{1/3}$MnO$_{3}$ (LSMO)
for the FM layers. We report the magnetic properties as functions
of temperature and spacer-layer thickness, using Polarized
Neutron Reflectivity (PNR) to map the magnetization profile
through the multilayer structure. Despite the structural
regularity of the interfaces, we find that magnetically disordered
regions appear at the surfaces of the LSMO layers. Contrastingly,
an enhanced FM moment is induced within the PCMO spacer when its
layer thickness is $\sim 3$ nm. Interestingly, this corresponds
to the situation where the layer thickness matches the size of
the FM clusters that occur in PCMO.

High quality epitaxial superlattices were grown on atomically
flat STO (001) substrates by pulsed laser deposition at 750
$^{\circ}$C. Then, the films were annealed at the same
temperature in 60 kPa of oxygen for 1 h. The multilayers involve
five repetitions of LSMO/PCMO bilayers, with LSMO as the starting
layer. We prepared seven different films, where the LSMO
thickness is in all cases 11.9 nm and the thickness of PCMO is
$t_A=0$, 0.8, 1.7, 2.7, 3.5, 4.3, and 7.6 nm, respectively (for
$t_A=0$, a single 11.9-nm-thick layer of LSMO was grown). The
out-of-plane lattice parameters (3.85 and 3.76 \AA$\;$for LSMO
and PCMO, respectively) are smaller than the bulk values (3.88
and 3.83 \AA, respectively), confirming the tensile stress
imposed by the substrate (the in-plane lattice parameter is 3.90
\AA$ $ for all STO, LSMO, and PCMO).

\begin{figure}
\begin{center}
\includegraphics[height=5.5cm,clip]{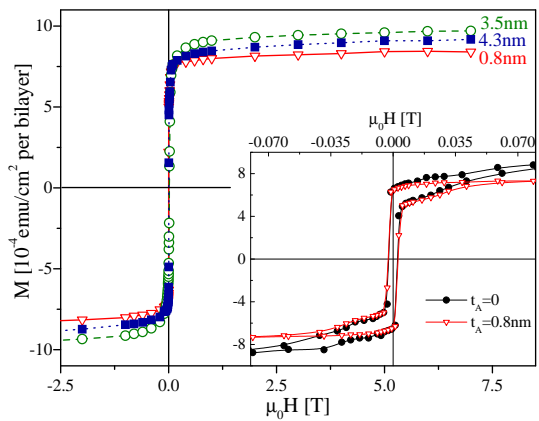}
\caption{(Color online) Magnetization loops at $T=5$ K for three
selected samples (1 emu = 10$^{-3}$ A m$^2$). The labels indicate
the thickness of the Pr$_{2/3}$Ca$_{1/3}$MnO$_{3}$ layers
($t_A$). Inset: low-field region of the same loops for $t_A=0$
and $0.8$ nm.} \label{Fig1}
\end{center}
\end{figure}

Magnetization ($M$) data were obtained in a superconducting
quantum interference device (SQUID) magnetometer, with in-plane
magnetic fields $\mu_0 H \leq 7$ T. $M(H)$ loops at a temperature
$T=5$ K for three selected samples are shown in Fig. 1. All our
multilayers show sharp FM loops, with a Curie temperature
$T_C\sim 345$ K (obtained from $M$ vs $T$) corresponding to the
LSMO layers. A close inspection to these data reveals a small but
clear variation of the magnetic moment with the thickness $t_A$.
After correcting for a minor substrate contribution \cite{STO},
we obtained the spontaneous magnetization of the samples ($M_0$,
the back-extrapolation to $H=0$ from the high-field saturated
region). Since only FM phases can give a spontaneous moment, $M_0$
is a direct measure of the FM volume of the samples
\cite{dario01}. Two curious features can be noticed in the $M_0$
vs $t_A$ curve [Fig. 2(a)]. At $t_A=0$, $M_0$ corresponds to the
saturation of pure LSMO, but then the moment decreases when a
thin layer of PCMO ($t_A=0.8$ nm) is added between the LSMO
layers. The decrease of $M_0$ suggests a reduction of the
magnetic moment within the LSMO layers, since any FM contribution
from the PCMO should align parallel to the applied field (this is
indeed confirmed by PNR). For thicker PCMO layers $M_0$ increases
again due to the contribution of FM droplets inside the PCMO
\cite{cheng98,dagotto01,smolyaninova02,radaelli01,saurel06}.

Among the interactions that could lead to the reduction of $M_0$,
an AFM coupling between different LSMO layers can be ruled out.
On one hand, RKKY-like interactions require a non-magnetic metal
as the intermediate layer \cite{RKKY}, which is not the case of
PCMO. On the other hand, the weak dipolar interactions could
hardly be responsible for this behavior, and the applied fields
should re-align the magnetic moments of the different LSMO
layers. However, $M$ stays the same even after cooling the
samples under a high field of 7 T. Differently, a plausible
reason for the reduction of $M_0$ is the formation of
magnetically disordered regions at the LSMO side of the FM-AFM
interfaces. Indeed, interface regions with degraded magnetization
have been observed in a number of different systems
\cite{TMR,hoffmann05,chakhalian06}.

\begin{figure}
\begin{center}
\includegraphics[height=8cm,clip]{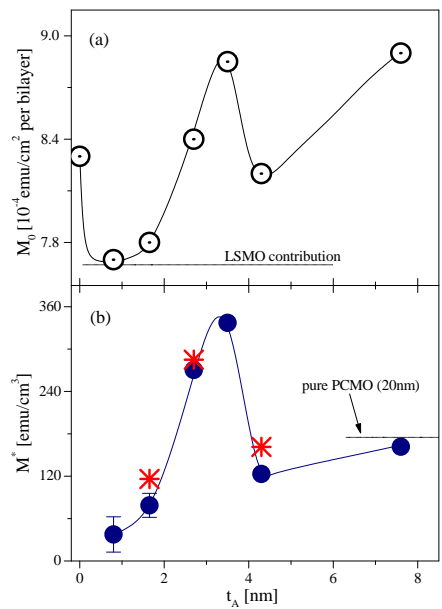}
\caption{(Color online) (a) Total FM moment of the samples at 5 K
as a function of $t_A$. The horizontal line is the estimated
contribution of the LSMO layers for $t_A>0$. (b) FM moment of the
Pr$_{2/3}$Ca$_{1/3}$MnO$_{3}$ layers only. The dotted line shows
the FM moment of a 20-nm-thick PCMO film. The FM moment obtained
from PNR at 6 K is also shown (star symbols).} \label{Fig2}
\end{center}
\end{figure}

While magnetization measurements cannot be conclusive about the
origin of the reduction of $M_0$, Polarized Neutron Reflectivity
provides a depth profile of $M$ and is thus an ideal tool to test
for the existence of disordered interfaces
\cite{hoffmann05,chakhalian06}. PNR experiments were carried out
on the NG1 reflectometer at the NIST Center for Neutron Research
(NCNR) at $T=6$, 120, and 300 K on three selected samples,
$t_A=1.7$, $2.7$, and $4.3$ nm. An in-plane magnetic field of 0.32
T was applied, enough to reach the saturation of the films (see
Fig. 1). Spin-flip scattering, which is indicative of
magnetization perpendicular to the applied field, was thus not
observed. As typical examples, the top panels of Fig. 3 show the
non-spin flip reflectivities ($R^{++}$ for spin up and $R^{--}$
for spin down) for $t_A=2.7$ nm at 300 K and for $t_A = 4.3$ nm
at 6 K (for the sake of clarity $R^{--}$ was multiplied by 10).
Note that the splitting between the non-spin flip reflectivities
is related to the projection of the magnetic moment parallel to
the applied field. Using the {\it Reflpak} software suite
\cite{reflpak}, the data were fitted to models for the depth
profile of the structure and magnetization after correcting for
polarization efficiencies ($> 97\%$) and other instrumental
effects. The models used to fit these data were kept as simple as
possible, while keeping a high quality fit. We must mention that
the statistical quality of the PNR data for $t_A=1.7$ nm is not
as good as for the other two samples studied, likely because of
the small thickness of PCMO. However, we note that these data
were successfully fitted with parameters that are consistent with
the magnetization results as well as with the other two samples
(see results below). The obtained $M$ profiles for $t_A=2.7$ and
$4.3$ nm are shown in the bottom panels of Fig. 3. For all three
samples, the fits at $T=300$ K unambiguously show the presence of
1.2-nm-thick regions on the LSMO side of the interfaces where $M$
is widely suppressed as compared to the inner part of the LSMO
layers, the so-called magnetically disordered interfaces (MDIs).
Although these MDIs are easily distinguished at $300$ K, at low
$T$ they do not show up very clearly. However, based on the drop
of $M_0$ at 5 K between the pure LSMO sample ($t_A=0$) and
$t_A=0.8$ nm [Fig. 2(a)], we estimate an effective thickness of
$\sim 0.5$ nm for these interface regions. This size is
comparable to the roughness of the interfaces and to the depth
resolution limits of PNR, which could mask the reduced
magnetization on this lengthscale.

\begin{figure}
\begin{center}
\includegraphics[height=7cm,clip]{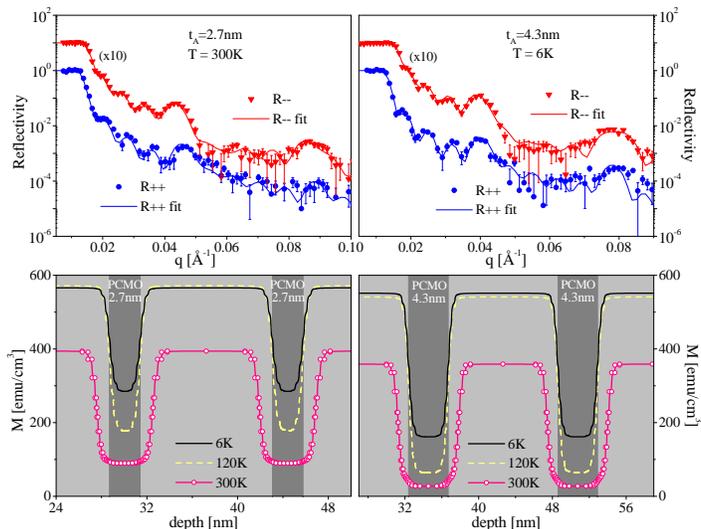}
\caption{(Color online) Top: Neutron reflectivity spectra for the
samples with $t_A=2.7$ nm (at $T=300$ K, left), and $t_A=4.3$ nm
(at $T=6$ K, right). The triangles and circles correspond to the
$R^{--}$ and $R^{++}$ reflectivity data, respectively ($R^{--}$
was multiplied by 10). The solid lines through the data are the
corresponding fits. Bottom: magnetization profiles at the three
studied temperatures (as labeled) for the same samples ($t_A=2.7$
and 4.3 nm on the left and right sides, respectively). The light
and dark gray areas are the LSMO and PCMO layers, respectively.}
\label{Fig3}
\end{center}
\end{figure}

MDIs were previously observed in several magnetic
heterostructures, including superlattices with FM-manganite layers
\cite{hoffmann05,chakhalian06}. However, more striking is the
presence of disordered regions at the free surfaces of manganite
samples \cite{park98,freeland05}. It has been argued that the
lower atomic coordination caused by the surface termination of the
crystal structure lowers the effective magnetic interactions,
leading to a degraded magnetization \cite{park98}. In
La$_{0.7}$Sr$_{0.3}$MnO$_{3}$ films, spin-resolved photoemission
spectroscopy (SPES) experiments show that, while at low $T$ full
saturation is reached, at higher $T$ (below $T_C$) several Mn-O
layers at the free surface of the film exhibit a reduced
magnetization as compared to the material underneath
\cite{park98}. This agrees with the general observation that the
MDIs are in fact regions where $M$ decays faster with $T$, and as
a consequence the TMR becomes deteriorated \cite{TMR}. Indeed,
also in our PNR experiments the MDIs are more clearly observed at
higher $T$ (see Fig. 3). Notwithstanding this, we believe that
even at low $T$ these disordered interface regions give rise to
anomalous features in our $M(H)$ measurements. As shown in the
inset of Fig. 1 for $t_A=0$ and $0.8$ nm, the reversal of $M$
when $H$ changes sign consists of two steps: first a fast
reversal occurs at $\mu_0 H<5$ mT, followed by a more gradual
moment increase at higher fields. For both samples the fast
reversal is similar, with the same values of $M$. However, at
higher fields the magnetization for $t_A=0.8$ nm stays clearly
lower than for $t_A=0$, indicating that a full saturation of $M$
is not reached. Since the difference between the two samples is
the contact of the LSMO surfaces with PCMO, this behavior must be
related to the interfaces. We conclude that the sharp reversal
corresponds to the inner volume of LSMO in the films, while the
gradual increase of $M$ is related to the alignment of the
disordered LSMO surfaces. At low $T$, full saturation can be
reached in the free LSMO surface in high magnetic fields, as
observed in the SPES experiments \cite{park98}. However, when the
PCMO layers are put in contact with the LSMO surfaces, MDIs
remain disordered in fields as large as 7 T and the magnetic
moment is smaller (inset of Fig. 1). We thus speculate that the
PCMO moments at the interfaces pin the magnetic moments of the
already disordered LSMO surfaces.

Due to the existence of these MDIs, the FM contribution of the
LSMO layers to the magnetic moment of the superlattices [dotted
line in Fig. 2(a)] is smaller than the saturation in the pure
LSMO film ($t_A=0$). This leads to the initial drop of $M_0$ at
low $t_A$ clearly seen in Fig. 2(a). As the thickness of PCMO
increases, the FM moment increases again and a second peculiar
feature develops, i.e. a maximum of $M_0$ at intermediate $t_A$.
It is well known that PCMO is a phase separated manganite, either
in bulk or thin film samples
\cite{smolyaninova02,radaelli01,saurel06}. Indeed, we also
prepared a 20-nm-thick PCMO film which exhibits spontaneous
ferromagnetism. Its FM moment $\sim 175$ emu/cm$^3$ appears
superposed to the linear $M$-$H$ response of the predominant AFM
background. In the multilayers, the FM moment of the PCMO layers
can be obtained after subtracting the contribution from the LSMO
[$\sim 7.7\times 10^{-4}$ emu/cm$^2$ each layer, dotted line in
Fig. 2(a)]. The obtained FM contribution of the PCMO layers
($M^*$) is plotted in Fig. 2(b) as a function of $t_A$.

The maximum of $M_0$ now appears as a maximum in $M^*$ vs $t_A$.
For very small $t_A$ values ($<2$ nm) the FM moment of PCMO
remains small, below 80 emu/cm$^3$. Presumably, the thin PCMO
layers are not large enough to accommodate FM clusters, and only
a small number of them can form. For large $t_A$, $M^*$
approaches the magnetic moment of the pure PCMO film (thickness
20 nm), as shown in Fig. 2(b). In between, the pronounced peak of
$M^*$ at $t_A$ around 3 nm is also confirmed by the PNR
experiments. When we compare the magnetic profiles of the samples
at $T=6$ K, the magnetization in the LSMO layers is observed to
be the same, and corresponds to the saturation of LSMO (solid
lines in the bottom panels of Fig. 3). The temperature dependence
of $M$ in the LSMO layers is also the same for all the samples
(square symbols in Fig. 4). However, the FM moment inside the
PCMO layers is clearly larger in the 2.7 nm sample. Moreover, as
demonstrated in Figs. 2(b) and 4, the values obtained from the PNR
fits at 6 K are in perfect agreement with those calculated from
$M(H)$ at 5 K. This maximum FM contribution of the PCMO layers may
originate from an overpopulation of FM clusters within the AFM
matrix. Of course, PNR does not show this {\it in-layer} nanoscale
structure since the depth-dependent $M$ profile represents an
average of the FM moment across the film plane.

\begin{figure}
\begin{center}
\includegraphics[height=5.5cm,clip]{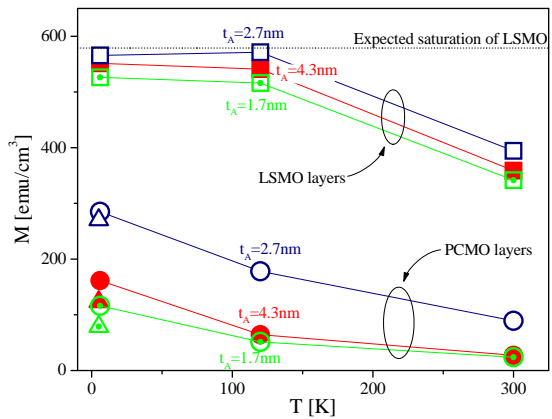}
\caption{(Color online) Temperature dependence of the FM moment
of the different layers of the LSMO/PCMO superlattices, as
obtained from PNR. The triangles are the magnetic moments of the
PCMO layers deduced from the magnetization measurements at $T=5$
K [$M^*$, see Fig. 2(b)].} \label{Fig4}
\end{center}
\end{figure}

The cause of such an increase of the FM moment of PCMO is not
obvious. The proximity of the FM LSMO layers could act as an
internal field \cite{krivorotov01,panagiotopoulos99}, raising the
number of FM droplets. This proximity effect would be related to a
strong influence of the FM layers on the AFM ones, mediated by
interface magnetic interactions \cite{chakhalian06}. However, as
we have already demonstrated the LSMO interfaces are magnetically
degraded, thus making difficult any kind of interlayer magnetic
coupling. This is especially clear at $T=300$ K, where the LSMO
layers show 1.2-nm-thick MDIs and the overshooting of the FM
moment in the PCMO for $t_A=2.7$ nm persists (Figs. 3 and 4). On
the other hand, strain effects could play a major role in this
behavior. Strain fields have been linked with nanoscale phase
separation in a number of materials, including manganite thin
films \cite{radaelli01,tao05}. In bulk PCMO samples the ground
state is a mainly charge-ordered (CO) AFM phase \cite{dagotto01}.
On the contrary, while thin films are still AFM the CO state is
not realized \cite{li02,krivorotov01,panagiotopoulos99,izumi00}.
Due to the strong electron-phonon coupling, strain fields play a
key role in the CO phase. Then, the absence of charge ordering is
likely related to the lattice strain and interface clamping in
thin film samples. In Pr$_{2/3}$Ca$_{1/3}$MnO$_{3}$, the
suppression of the CO phase could make the material especially
unstable against the formation of FM clusters. Indeed, it is
highly unusual that the FM contribution in the PCMO layers
persists even at high $T$ (300 K), as shown in Figs. 3 and 4. On
the contrary, at this $T$ a FM response is totally absent in bulk
samples, confirming that strain fields must play a fundamental
role in the stabilization of FM nanoclusters in our films.
Likely, a matching condition between layer thickness and cluster
size is responsible for the maximum FM moment of the PCMO layers
at $t_A\sim 3$ nm. Indeed, neutron scattering experiments show a
similar lengthscale for the typical FM cluster size
\cite{saurel06}. The accommodation of nanometric droplets in a
strain field occurring at a similar lengthscale would favor an
overpopulation of FM clusters.

In summary, the combination of PNR and magnetometry allowed us to
reveal the nanoscale magnetic structure of the LSMO/PCMO
multilayers. The stacking of PCMO on top of LSMO hampers the
saturation at the surfaces of the ferromagnetic LSMO layers, which
show 1.2-nm-thick magnetically disordered interfaces at 300 K. In
the phase separated PCMO, a maximum FM moment occurs for layer
thicknesses comparable to the characteristic lengthscale of the
FM nanoclusters, i.e. for $t_A\sim 3$ nm. This enhancement of
ferromagnetism in the nominally AFM spacer could have important
implications for the TMR response of these devices. In similar
manganite multilayers an enhanced TMR was shown to be related to
a partial metallization of the insulating layers at intermediate
optimal thicknesses \cite{cheng98,li02}. In our superlattices,
the maximum FM moment would also imply a tendency towards the
metallization of the PCMO spacer, achieved through the specific
control of strain.


\begin{thebibliography}
\bibitem{STO} A tiny magnetic contribution was found in
the clean STO substrates. This could be due to 1-2 ppm of Fe
impurities that are present even in the best STO crystals
available.
\end{thebibliography}
\end{document}